\documentclass[11pt,journal,draftcls,letterpaper,onecolumn]{IEEEtran}

\usepackage{graphicx} 
\usepackage{amsmath}
\usepackage{amssymb}
\newcommand{\comment}[1]{}

\newcommand{\T}{^{\scriptscriptstyle \top}}
\newcommand{\HH}{\mathbf H}
\newcommand{\re}{\mbox{Re}}
\newcommand{\im}{\mbox{Im}}

\newcommand{\mH}{\hbox{{\bf H}}}
\newcommand{\vn}{\mbox{${\epsilon}$}}
\newcommand{\vh}{\mbox{${\bf h}$}}
\newcommand{\bs}{\mbox{${x}$}}

\newcommand{\vx}{y}

\newcommand{\vs}{\mbox{${x}$}}
\newcommand{\mC}{\mathbf C}
\newcommand{\vz}{z}
\newcommand{\beq}{\begin{equation}}
\newcommand{\eeq}{\end{equation}}
\newcommand{\bea}{\begin{array}}
\newcommand{\ena}{\end{array}}

\begin{document}

 \title{MIMO Detection for High-Order QAM Based on a Gaussian Tree Approximation}

\author{Jacob Goldberger
        \thanks{J. Goldberger is with the School of Engineering,
        Bar-Ilan University, 52900, Ramat-Gan. Email: goldbej@eng.biu.ac.il}
        and Amir Leshem \thanks{A. Leshem is with the School of Engineering,
        Bar-Ilan University, 52900, Ramat-Gan. Email: leshema@eng.biu.ac.il}
   }
\date{}
 \maketitle

\begin{abstract}
This paper proposes a new detection algorithm for MIMO communication systems employing high order QAM constellations.
 The factor graph that corresponds to this problem is
very loopy; in fact, it is a complete graph. Hence, a straightforward application of the
Belief Propagation (BP) algorithm yields very poor results.
Our algorithm is based on  an optimal  tree approximation of the Gaussian density of the unconstrained linear system.
The finite-set constraint is then applied to obtain a loop-free discrete distribution.
It is shown that even though the approximation
is not directly applied to the exact discrete distribution,
applying the BP algorithm to the loop-free factor graph
outperforms current methods in terms of both performance and complexity.
The improved performance of the proposed
algorithm is demonstrated  on the problem of MIMO detection.
\end{abstract}

\comment{}
\section{Introduction}
\label{sec:intro}

Finding a linear least squares fit to data is a  well-known problem,
with applications in almost every field of science. When there are
no restrictions on the variables, the problem has a closed form
solution. In many cases, a-priori knowledge on the values of the
variables is available. One example is the existence of priors,
which leads to Bayesian estimators. Another example  of great
interest in a variety of areas is when the variables are
constrained  to a discrete finite set.
This problem has many diverse applications such as the decoding of
multi-input-multi-output (MIMO) digital communication systems. In contrast to the continuous linear least
squares problem, this problem is known to be NP hard \cite{Fincke}.

We consider a MIMO communication system with
$n$ transmit antennas and $m$ receive antennas. The
 tap gain from transmit antenna $i$ to receive
antenna $j$ is denoted by $\HH_{ij}$. In each use of the MIMO
channel a  vector $x=(x_1,...,x_n)^{\T}$ is independently selected
from a finite set of points $\mathcal{A}$ according to the data to
be transmitted, so that $x \in \mathcal{A}^n$.
 The received vector $y$ is given by:
\begin{equation} y = \HH x + \epsilon  \label{model}
\end{equation}
The vector $\epsilon$ is an additive noise in which the noise
components are assumed to be zero mean, statistically independent
Gaussians with a known variance $\sigma^2 I$. The $m\!\times\!n$
matrix $\HH$ is comprises i.i.d. elements drawn from a
complex  normal distribution of unit
variance.  The MIMO detection problem consists of finding the
unknown transmitted vector $x$ given $\HH$ and $y$. The task, therefore, boils down to solving a linear system  in which the unknowns are
constrained to a discrete finite set.
It is convenient to reformulate the  complex-valued model into a real valued one. It can be translated into an
equivalent double-size real-valued representation that is obtained
by considering the real and imaginary parts separately:
\begin{equation} \nonumber
 \left[\!
         \begin{array}{c}
           \re(\vx) \\ \im(\vx)
         \end{array} \!
       \right] = \left[\! \begin{array}{cc}
                   \re(\mH) & -\im(\mH) \\
                   \im(\mH) & \hspace{0.2cm}\re(\mH)
                 \end{array} \! \right]
                        \left[\! \begin{array}{c}
           \re(\bs) \\ \im(\bs) \end{array}  \!\right] +
\left[ \! \begin{array}{c}
           \re(\vn) \\ \im(\vn)
         \end{array} \! \right]
\end{equation}
Hence we assume hereafter that $\mH$ has real values without any
loss of generality. The maximum
likelihood (ML) solution is:
\begin{equation}
\hat{x} = \arg \min_{x\in \mathcal{A}^n} \| \HH x - y\|^2
\end{equation}
However, going over all the $|\mathcal{A}|^n$ vectors is unfeasible when either $n$ or $|\mathcal{A}|$ are large.

 A simple sub-optimal solution, known as the Zero-Forcing algorithm, is based on a linear decision that ignores the
finite set constraint:
\begin{equation}
\label{zf} z = (\HH^{\T}\! \HH )^{-1} \HH^{\T}\! y
\end{equation}
 and then, neglecting the correlation between the symbols,
 finding the closest  point in $\mathcal{A}$ for each symbol
independently:
\begin{equation}
\hat{x}_i = \arg \min_{a\in \mathcal{A}} | z_i - a | \label{hzf}
\end{equation}
 This scheme performs poorly due to its inability to
handle ill-conditioned realizations of the matrix $\HH$. Somewhat
better performance can be obtained by using a minimum mean square
error (MMSE) Bayesian estimation  on the continuous linear
system. Let $e$ be  the mean symbol energy.
We can partially incorporate the information that $x\in \mathcal{A}^n$ by using the prior Gaussian
distribution $x \sim \mathcal{N} ( 0 , e I)$. The MMSE estimation becomes:
\begin{equation}
\label{mmse}  E(x|y) = (\HH^{\T}\! \HH + \frac{\sigma^2}{e}
I )^{-1} \HH^{\T}\! y
\end{equation}
and then  the finite-set solution is obtained by finding the closest
lattice point in each component independently. A vast improvement
over the linear approaches described above can be achieved by the V-BLAST algorithm that is
based on sequential decoding with optimal ordering \cite{Golden}.

These linear type algorithms
can also easily provide probabilistic (soft-decision) estimates for
each symbol. However, there is  still a significant gap between the
detection performance of the V-BLAST algorithm and the performance
of the  ML detector.

Many alternative methods have been proposed to approach the ML
detection performance. The  sphere decoding (SD) algorithm finds the exact ML solution by searching the nearest lattice point. \cite{Fincke,Schnorr,Boutros,Studer}.
     Although the SD reduces
computational complexity compared to the exhaustive search of ML solution, sphere decoding is not feasible for high-order QAM constellations.  While sphere decoding has been empirically found to be computationally very fast for small to moderate problem sizes (say, for $n<20$ for 16-QAM), the sphere decoding complexity would be prohibitive for large $n$, higher order QAM  and/or low SNRs \cite{xJalden2005}.
Another family of MIMO decoding algorithms is based on  semidefinite relaxation
(e.g. \cite{Weisel,Sidiropoulos,Ma}). Although the theoretical computational complexity of semidefinite relaxation is a low degree polynomial, in practice the running time is very high.  Thus, there  is still a need for low complexity detection algorithms that perform well.

This study attempts to solve the MIMO decoding problem
  using the Belief Propagation (BP) paradigm. It is well-known  (see e.g. \cite{Shental2008}) that a straightforward
implementation of the BP algorithm to the MIMO detection  problem
yields very poor results
 since there are a large number of short cycles in the underlying factor graph.
  In this study we introduce a novel approach to utilize the BP paradigm for
MIMO detection. The proposed variant of the BP algorithm is both
computationally efficient and achieves near optimal results.
A preliminary version of this paper appears in \cite{Goldberger_NIPS}.
 The paper
proceeds as follows. In Section II we discuss previous attempts to apply variants of the BP algorithm to the MIMO decoding problem.
The proposed algorithm which we dub `The Gaussian-Tree-Approximation (GTA)  Algorithm'
is described in Section III. Experimental results are presented in Section IV.

\section{The Loopy Belief Propagation Approach}
\begin{figure}[th]
       \centering
         \includegraphics[keepaspectratio=true,clip=true, angle=0,
totalheight=0.28\textheight] {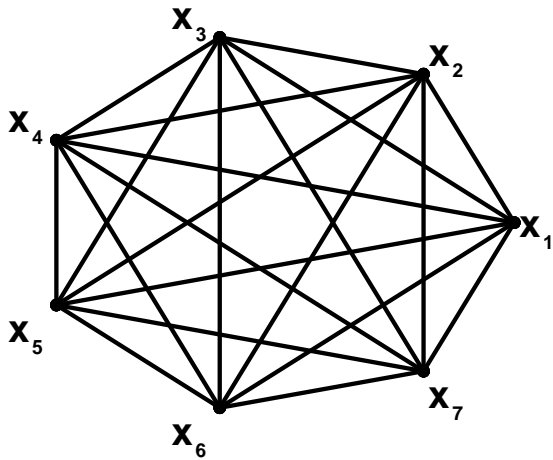} \caption{The MRF undirected  graphical model corresponds to  the MIMO detection problem with $n=7$.}
\label{mf_graph}
\end{figure}

Given the constrained linear system  $y = \HH x + \epsilon$, and a uniform
prior distribution on $x$, the posterior probability function of the discrete random vector $x$ given $y$ is:
\begin{equation}
p(x|y)   \propto \mbox{exp} ( -\frac{1}{2 \sigma^2} \|\HH x-y\|^2)  \hspace{1cm},\hspace{1cm}  x\in \mathcal{A}^n
\label{pxy}
\end{equation}
The notation $\propto$ stands for equality up to a normalization
constant. Observing that $\|\HH x-y\|^2$ is a quadratic expression, it can be easily verified that $p(x|y)$ is factorized into
a product of two-  and single-variable potentials:
\begin{equation}
p(x_1,..,x_n|y) \propto \prod_i {\psi_i(x_i)} \prod_{i<j}
{\psi_{ij}(x_i,x_j)} \label{pfactors}
\end{equation}
such that \begin{equation}  \psi_i(x_i)  =  \exp(-\frac{1}{2
\sigma^2} y^{\T}\vh_i x_i)\end{equation}
$$ \psi_{ij}(x_i,x_j)  =  \exp(-\frac{1}{\sigma^2}\vh_i^{\T}\vh_j x_i
x_j)$$
 where $\vh_i$ is the $i$-th column of the matrix $\HH$.
Since the obtained factors are simply a function of pairs, we obtain a
Markov Random Field (MRF) representation \cite{Yedidia}.  In the MIMO application the
(known) matrix $\HH$ is randomly selected and therefore, the MRF graph is usually a completely connected graph (see an MRF graph illustration in Fig. \ref{mf_graph}).

The Belief Propagation (BP) algorithm aims to solve inference problems
by propagating information throughout this MRF via a series of messages
sent between neighboring nodes (see \cite{Sudderth} for an excellent tutorial on BP). In the sum-product variant of the BP algorithm applied to the MRF  (\ref{pfactors}), the message from
$x_j$ to $x_i$ is:
\begin{equation}
\label{bpiter} m_{j\rightarrow i}(x_i) = \sum_{x_j\in \mathcal{A}}
(\psi_j(x_j)  \psi_{ij}(x_i,x_j)  \prod_{k\ne i,j } m_{k\rightarrow
j}(x_j)) \hspace{0.2cm},\hspace{0.2cm}  x_i\in \mathcal{A}
\end{equation}
In each  iteration messages are passed along all the graph edges in
both edge directions.  In every iteration, an estimate of the
posterior marginal distribution (`belief') for each variable can be
computed by multiplying together all of the incoming messages from
all the other nodes: \begin{equation}b_i(x_i) = \psi_i(x_i)  \prod_{k\ne i}
m_{k\rightarrow i}(x_i) \hspace{1cm},\hspace{1cm}  x_i\in \mathcal{A}\end{equation} A variant of the sum-product algorithm is the max-product algorithm in
which the summation in Eq. (\ref{bpiter}) is replaced by a
maximization over all the symbols in $\mathcal{A}$. In a loop-free
MRF graph the sum-product algorithm always converges to the exact
marginal probabilities (which corresponds in the case of MIMO
detection to a soft decision probability of each symbol $p(x_i|y)$).
  In a loop-free MRF graph the max-product variant of the BP  algorithm always converges to the   most likely configuration \cite{Pearl} (which corresponds to ML decoding
in our case). For loop-free graphs, BP is essentially a distributed
variant of dynamic programming. The BP message update equations only
involve passing messages between neighboring nodes. Computationally,
it is thus straightforward to apply the same local message updates
in graphs with cycles. In most such models, however, this loopy BP
algorithm will not compute exact marginal distributions; hence,
there is almost no theoretical justification for applying the BP
algorithm (one exception is that, for Gaussian graphs, if BP
converges, then the means are correct \cite{Weiss2001}).
 However, the BP algorithm applied to
loopy graphs  has been found to have outstanding empirical success
in many applications, e.g., in decoding  LDPC codes \cite{Gallager}.
The performance of BP in this application may be attributed to the
sparsity of the graphs. The cycles in the graph are long, hence the
graph has tree-like properties, so that messages are approximately
independent and inference may be performed as though the graph was
loop-free.
 The BP algorithm  has also been
used successfully  in  image processing and computer vision (e.g.
\cite{Felzenszwalb}) where the image is represented by a
grid-structured MRF that is based on local connections between
neighboring nodes.

\begin{figure}[th]
       \centering
         \includegraphics[keepaspectratio=true,clip=true, angle=0,
totalheight=0.22\textheight] {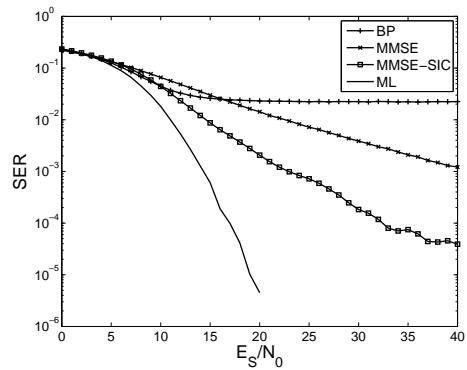} \caption{Decoding
results for $8\times 8$ BPSK real valued system, $\mathcal{A}=\{-1,1\}$.}
\label{bporig}
\end{figure}

However, when the graph is not sparse, and is not based on local
grid connections,  loopy BP almost always fails to converge. Unlike
the sparse graphs of LDPC codes, or grid graphs in computer vision
applications, the MRF graphs of MIMO channels are  {\it completely
connected graphs} and therefore the associated detection performance
is poor. This has prevented the BP from being an asset for the MIMO
problem. Fig. \ref{bporig} shows an example of a BPSK MIMO
system based on an $8\times 8$ matrix and $\mathcal{A}=\{-1,1\}$
(see Section IV for a detailed description of the
simulation set-up). As can be seen in Fig. \ref{bporig}, the BP
decoder  based on the MRF representation (\ref{pfactors}) has
very poor results. Standard techniques to stabilize the BP iterations such as damping  the message updates \cite{Murphy} do not help here. Even applying more advanced versions of BP (e.g. Generalized BP and Expectation Propagation) to inference problems on complete MRF graphs yields poor results \cite{Minka}. The problem here is not in the optimization method but in the cost function that needs to be modified to yield a good approximate solution.

\begin{figure}[th]
       \centering
         \includegraphics[keepaspectratio=true,clip=true, angle=0,
totalheight=0.22\textheight] {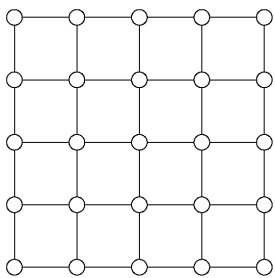} \caption{The MRF grid model corresponds to $5 \times 5$ 2D intersymbol interference (ISI) channels.}
\label{grid}
\end{figure}

 There have been several recent
 attempts to apply BP to the MIMO detection problem with
good results (e.g. \cite{Kabashima,Neirotti,Hu,Kaynak}). However,  in these methods
the factorization of the
probability function is done in such a way that each factor
corresponds to a single linear equation. This leads to a partition
of the probability function into factors each of which is a function
of all the unknown variables. This results in an exponential
computational complexity when  computing the BP messages. Shental et.
al \cite{Shental2008} analyzed the case where the matrix $\HH$ is
relatively sparse (and has a grid structure) (see Fig. \ref{grid}). They showed that even
under this restricted assumption the BP still does not perform well.
As an alternative method they proposed the generalized belief
propagation (GBP) algorithm that does work well on the sparse matrix
if the algorithm regions are carefully chosen. There are situations
where the sparsity assumption makes sense (e.g. 2D intersymbol
interference (ISI) channels). However, in the MIMO channel model we
assume that the channel matrix elements are i.i.d. and Gaussian; hence
we cannot assume that the channel matrix $\HH$ is sparse.

In the recent years there were also several attempts to apply BP for densely
connected graphs (mainly Gaussian graphs) \cite{Montanari,Bickson,Bayati,Kabashima,Neirotti}.

\section{The Gaussian Tree Approximation Algorithm}

Our approach is based on an approximation of the exact probability
function:
\begin{equation}
 p(x_1,..,x_n|y) \propto \mbox{exp} ( -\frac{1}{2 \sigma^2}
\|\HH x-y\|^2) \hspace{0.5cm},\hspace{0.5cm}  x\in \mathcal{A}^n \label{exact_ml}\end{equation} that enables a successful implementation of the Belief Propagation paradigm. Since the BP algorithm is optimal on loop-free factor graphs (trees) a reasonable approach is finding an optimal tree approximation of the exact distribution (\ref{exact_ml}).
Chow and Liu \cite{Chow} proposed a method to find a tree approximation of a given distribution that has the minimal Kullback-Leibler (KL) divergence to the true distribution. They showed that the optimal tree  can be learned efficiently via a maximum spanning tree whose edge weights correspond to the mutual information between the two variables corresponding to the edge's endpoints. The problem is that the Chow-Liu algorithm is based on  the two-dimensional marginal distributions. However, finding the marginal distribution of the probability function (\ref{exact_ml}) is, unfortunately, NP hard and it is (equivalent to)  our final target.

To overcome this obstacle, our approach is based on applying  the  Chow-Liu algorithm on the distribution corresponding to the unconstrained linear system. This distribution is Gaussian and therefore it is straightforward in this case to compute the two-dimensional marginal distributions. Given the Gaussian tree approximation, the next step of our approach  is to apply the finite-set constraint and utilize the Gaussian tree distribution to form a discrete loop free approximation of $p(x|y)$  which can be efficiently globally maximized using the BP algorithm. To motivate this approach we first apply a simplified version  to derive the zero-forcing decoding algorithm  (\ref{hzf}) described in Section I.

Let $z(y)=(\HH^{\T} \HH )^{-1}\HH^{\T}\!y$ be  the
 least-squares  estimator (\ref{zf})  and
$\mC=\sigma^2(\HH^{\T} \HH )^{-1}$ is its variance. It can be easily verified that $p(x|y)$ (\ref{exact_ml}) can be written as:
\begin{equation}
p(x|y) \propto f(x;z,C) = \frac{1}{\sqrt{(2\pi)^n |C|}}\exp(-\frac{1}{2}(
z-x)^{\T}C^{-1}(z-x)) \label{pxyfzf} \end{equation}
where $f(x;z,C)$ is a Gaussian density with mean $z$ and covariance matrix $C$.
$f(x;z,C)$ can be viewed as a posterior distribution of $x$ assuming a non-informative prior. Now, instead of
marginalizing the true distribution $p(x|y)$, which is an NP hard problem,  we approximate it by the product of the  marginals of the Gaussian density $f(x;z,C)$:
\begin{equation} f(x;z,C) \approx \prod_i f(x_i;z_i,C_{ii}) = \prod_i\frac{1}{\sqrt{2\pi C_{ii}}}\exp( -\frac{(z_i-x_i)^2}{2C_{ii}})
\label{1dappr}\end{equation}
At this stage we apply the finite-set constraint.
 From the
Gaussian approximation (\ref{1dappr}) we can extract a discrete approximation:
\begin{equation}
\hat p(x|y) \propto \prod_i \exp( -\frac{(z_i-x_i)^2}{2\mC_{ii}}) \hspace{0.5cm}
, \hspace{0.5cm} x\in \mathcal{A}^n
\end{equation}
Since this joint probability function is obtained as a product of marginal probabilities,
we can address each variable separately:
\begin{equation}
\hat{p}(x_i=a|y) \propto  \exp ( -\frac{(z_i-a)^2}{2\mC_{ii}}) \hspace{0.5cm}
, \hspace{0.5cm} a\in \mathcal{A}
\end{equation}
Taking the most likely symbol we obtain the sub-optimal Zero-Forcing solution (\ref{hzf}).

Motivated by the simple product-of-marginals approximation  described above, we suggest  approximating the discrete distribution $p(x|y)$ via a tree-based approximation of the Gaussian distribution $f(x;z,C)$. Although the Chow-Liu algorithm was originally stated for discrete distributions, one can  verify that it also applies for the Gaussian case.
For the sake of completeness we give a  detailed derivation below.

\subsection{Finding the optimal Gaussian tree approximation}

\newcommand{\xip}{(x_i|x_{p(i)})}
\newcommand{\xjp}{(x_i,x_{p(i)})}

 We represent an $n$-node tree graph by the loop-free parent relations $\{p(i)\}_{i=1}^n$
  such that $p(i)$ is the parent node of $i$.
 To simplify notation we do not separately describe the root node. The  parent of the root is  implicitly assumed to be the  empty set. A distribution $g(x_1,...,x_n)$ is described by a tree $\{p(i)\}$ if it can be written as  $g(x) = \prod_{i=1}^n g \xip $.
 We start with a formula for the KL divergence between a Gaussian distribution $f(x)$ and a distribution $g(x)$ defined on the same space that is represented by
a tree graphical model.

\noindent {\bf Theorem 1:} Let $f(x) = f(x_1,...,x_n)$ be a multivariate Gaussian distribution and let
$g(x) = \prod_{i=1}^n g \xip $ be another distribution that is represented by a loop-free graphical model (tree). The KL divergence between $f$ and $g$ is:
\begin{eqnarray}
D( f||g)  & = &  \sum_{i=1}^n  D ( f \xip || g \xip )
\label{dfg} \\
 & & -h(x) +\sum_{i=1}^n ( h(x_i) - I ( x_i ; x_{p(i)})) \nonumber
\end{eqnarray}
such that $I$ is the mutual information and $h$ is the differential entropy, based on the distribution $f(x)$.

\noindent {\bf Proof:} The definition of the KL divergence implies:
\begin{eqnarray*}
D(f||g) & = &  \int f \log f - \int f(x)  \sum_{i=1}^n \log g \xip \\
  & = & -h(x) - \sum_{i=1}^n \int f \xjp  \log g \xip  \\
 & = & -h(x) + \sum_{i=1}^n \int f \xjp  \log \frac{f \xip}{g \xip } \\
  & & - \sum_{i=1}^n \int f \xjp  \log f \xip
\end{eqnarray*}
 \begin{eqnarray*}
 &  = & -h(x) + \sum_{i=1}^n  D ( f \xip || g \xip )\\ & &  - \sum_{i=1}^n  \left( I ( x_i ; x_{p(i)}) - h(x_i) \right)
 \hspace{2.0cm} \Box  \end{eqnarray*}

From Eq. (\ref{dfg}) it can be easily seen that if we fix  a tree graph $\{p(i)\}_{i=1}^n$, the tree distribution whose KL divergence to $f(x)$ is minimal is $g(x) = \prod_{i=1}^n f \xip$. In other words, the best tree approximation of  $f(x)$ is constructed from the conditional distributions of $f(x)$. For that tree approximation we have:
\begin{equation}
D( f||\prod_{i=1}^n f \xip) = -h(x) + \sum_{i=1}^n  h(x_i) - \sum_{i=1}^n  I ( x_i ; x_{p(i)})
\label{dftf}
\end{equation}
Moreover, since in Eq. (\ref{dftf}) $h(x)$ and $h(x_i)$ do not depend on the tree structure,  the tree topology $\{p(i)\}$ that best approximates $f(x)$ is the one that maximizes the sum: \begin{equation}\sum_{i=1}^n I( x_i ; x_{p(i)}) \label{summi}\end{equation}
A spanning tree of a graph is a subgraph that contains all the vertices and is a tree.
Now suppose the edges of the graph have weights. The weight of a spanning tree is simply the sum of weights of its edges.  Eq. (\ref{summi}) reveals that the problem of finding the best tree approximation of the Gaussian distribution $f(x)$ can be reduced to the well-known problem of finding the maximum spanning tree of the weighted $n$-node graph where the weight of the $i$-$j$ edge is the mutual information between $x_i$ and $x_j$ \cite{Chow}. It can be easily verified that the mutual information between two r.v. $x_i$ and $x_j$ that are jointly Gaussian is:
\begin{equation}I(x_i;x_j) =  -\log ( 1-\rho_{ij}^2) \end{equation}
where $\rho_{ij}$ is the correlation coefficient between $x_i$ and $x_j$.

There are several algorithms to find a minimum spanning tree. They  all utilize a greedy approach. In this work we use the Prim algorithm \cite{Prim} which is  efficient and very simple to implement. The Prim algorithm begins with some vertex $v$ in a given graph, defining the initial set of vertices $T$. Then, in each iteration, we choose a minimum-weight edge $(u,v)$, connecting a vertex $v$ in the set $T$ to the vertex $u$ outside of set $T$. Then vertex $u$ is brought in to $T$. This process is repeated until a spanning tree is formed.  We can use a heap to remember, for each vertex, the smallest edge connecting the current sub-tree $T$ with that vertex.  The complexity of the Prim's algorithm, that finds the minimum spanning tree of an $n$-vertex  graph is $O(n^2)$.

We note in passing that since the Prim algorithm is based on a greedy approach,
it only relies on the order of the weights and not on their exact values.
Hence, applying a monotonically increasing function on the graph weights does not change the topology of the optimal tree. To find the optimal Gaussian tree approximation we can, therefore,  use the weights $\rho_{ij}^2$ instead of $I(x_i;x_j) = -\log ( 1-\rho_{ij}^2)$.
  The optimal Gaussian tree is, therefore,  the one that maximizes the  sum of the square correlation coefficients between adjacent nodes.
To summarize, the algorithm that finds the best Gaussian tree approximation is as follows.  Define $x_1$ as the root. Then find the edge connecting a vertex in the tree to the vertex outside the tree, such that the corresponding  square correlation coefficient is maximal and add the edge to the tree. Continue this procedure until a spanning tree is obtained.

\subsection{Applying BP on the tree approximation}

Let $\hat{f}(x)$ be the optimal Chow-Liu tree approximation of $f(x;z,C)$ (\ref{pxyfzf}). We can assume, without loss of generality, that $\hat{f}(x)$ is rooted at $x_1$.
$\hat{f}(x)$ is a loop-free Gaussian distribution on $x_1,...,x_n$, i.e.
 \begin{equation}\hat{f}(x) = f(x_1;z,C) \prod_{i=2}^n f(x_i | x_{p(i)};z,C) \hspace{0.3cm},\hspace{0.3cm} x\in R^n
 \end{equation}
where $p(i)$ is the `parent' of the $i$-th node in the optimal tree.
The Chow-Liu algorithm guarantees that $\hat{f}(x)$ is the optimal Gaussian tree approximation of $f(x;z,C)$ in the sense that the KL divergence $D(f||\hat{f})$ is minimal.

 Given the Gaussian tree approximation, the next step of our approach  is to apply the finite-set constraint  to form a discrete loop free approximation of $p(x|y)$  which can be efficiently globally maximized using the BP algorithm.
Our approximation approach is, therefore, based on replacing the
true distribution $p(x|y)$ with the following approximation:
\begin{equation}
\hat{p}(x_1,...,x_n|y) \propto  \hat{f}(x) = f(x_1;z,C) \prod_{i=2}^n f(x_i | x_{p(i)};z,C)
\hspace{1cm},\hspace{1cm} x\in \mathcal{A}^n \label{appr}
\end{equation}
\comment{Note that comparing with the MMSE method there are two we utilize the Gaussian distribution $f(x,z;C)$ to find a markovian approximation both the tree topology of the discrete approximation and to find the parameters of the discrete Markovian distribution derived from the tree. We show in the experiment section that both features are important to obtain an improved algorithm.}

The probability function $\hat{p}(x|y)$ is a loop free factor graph.
Hence the BP algorithm can be applied to find its most likely configuration.
We next derive the messages  of the BP algorithm that is applied on  $\hat{p}(x|y)$.
An optimal BP schedule, when applied to a tree, requires passing a message  once in each direction of each edge \cite{Kschischang}. The BP messages are first sent from leaf variables `downward' to the root.
The computation begins at the leaves of the graph. Each
leaf variable node sends a  message to
its parent.  Each vertex waits for messages from all of its
children before computing the message to be sent to its parent.
 The `downward' BP message from a variable $x_i$ to its parent variable $x_{p(i)}$ is computed based on all the messages $x_i$ received from its children:
 \begin{equation}
 m_{i\rightarrow p(i)}(x_{p(i)}) = \sum_{x_i\in \mathcal{A}}  f(x_i | x_{p(i)};z,C) \prod_{j|p(j)=i}   m_{j\rightarrow i }(x_i) \label{bpmup}
 \end{equation}
 If $x_i$ is a leaf node in the tree then the message is simply:
 \begin{equation}
 m_{i\rightarrow p(i)}(x_{p(i)}) = \sum_{x_i\in \mathcal{A}}  f(x_i | x_{p(i)};z,C)  \end{equation}
The `downward' computation terminates at the root node.

\begin{figure*}
\vspace{0.2cm} {
  {\centering \fbox{ \centering
\parbox{6.0in}{

  {Input:} A constrained linear LS problem: $\HH \vs + \vn = \vx$, a noise
   level $\sigma^2$ and a finite symbol set $\mathcal{A}$ whose the mean symbol energy is denoted by $e$.

\vspace{0.3cm}

{Algorithm:}

\begin{itemize}
\item Compute $\vz=(\HH^{\T}\HH + \frac{\sigma^2}{e}I)^{-1}\HH^{\T} \! y$ and
$\mC=\sigma^2(\HH^{\T}\HH+ \frac{\sigma^2}{e}I )^{-1}$.

\item Denote: \begin{eqnarray*}
 f(x_i;z,C) & = &    \exp ( -\frac{1}{2}\frac{(x_i-z_i)^{2}}{C_{ii}}) \\
f(x_i|x_j;z,C) & = &  \exp( -\frac{1}{2} \frac{((x_i-z_i)-C_{ij}/C_{jj}(x_{j}-z_{j}))^2}
{C_{ii}-C_{ij}^2/C_{jj}})
\end{eqnarray*}

\item Compute maximum spanning tree of the $n$-node  graph where the weight of the $i$-$j$ edge is the square of the correlation coefficient: $ \rho_{ij}^2 = C_{ij}^2/(C_{ii}C_{jj})$

Assume the tree is rooted at node `1' and denote the parent of node $i$ by $p(i)$.

\item Apply BP on the loop free distribution:
$$\hat{p}(x_1,...,x_n|y) \propto  f(x_1;z,C) \prod_{i=2}^n f(x_i | x_{p(i)};z,C)
\hspace{1cm} x_1,...,x_n\in \mathcal{A} $$
to find the (approx. to the)  most likely configuration.
\end{itemize}
    }}}}
    \label{alg_box}
    \caption{The Gaussian Tree Approximation (GTA) Algorithm.}
    \end{figure*}

Next, BP messages are sent `upward' back to the leaves.
The computation begins at the root of the graph. Each
 vertex waits for a message from its parent
before computing the messages to be sent to each of its children.
 The `upward' BP message from a parent variable $x_{p(i)}$ to its child variable $x_i$ is computed based on the `upward' message  $x_{p(i)}$ received from its parent  $x_{p(p(i))}$  and from `downward' messages that $x_{p(i)}$  received from all the siblings of $x_i$:
    \begin{equation}
 m_{p(i)\rightarrow i}(x_i) = \sum_{x_{p(i)}\in \mathcal{A}}  f(x_i | x_{p(i)};z,C)  m_{p(p(i))\rightarrow p(i)} (x_{p(i)}) \times \label{bpmdown}
 \end{equation}
$$  \prod_{\{j|j\ne i,p(j)=p(i)\}}   m_{j\rightarrow p(i) }(x_{p(i)}) \hspace{0.6cm},\hspace{0.6cm} x_i\in \mathcal{A} $$
If $x_{p(i)}$ is the root of  the tree then the message is simply:
    \begin{equation}
 m_{p(i)\rightarrow i}(x_i) = \sum_{x_{p(i)}\in \mathcal{A}}  f(x_i , x_{p(i)};z,C)  \times
 \label{bpmrootdown}
 \end{equation}
$$ \prod_{\{j|j\ne i,p(j)=p(i)\}}  m_{j\rightarrow p(i) }(x_{p(i)}) \hspace{1cm},\hspace{1cm} x_i\in \mathcal{A}$$

 After the downward-upward message passing procedure is completed we can compute the `belief'
  at each variable which is the product of all the messages sent to the variable from its parent and from its children (if there are any).
     \begin{equation}
 \mbox{belief}_i(x_i) = m_{p(i)\rightarrow i} (x_i) \prod_{j|p(j)=i}   m_{j\rightarrow i }(x_i)  \hspace{0.2cm},\hspace{0.2cm} x_i\in \mathcal{A} \label{bpmbel}
 \end{equation}
 In the case $x_i$ is the root node,  the `belief'  computed is as follows:
    \begin{equation}
 \mbox{belief}_i(x_i) = f(x_i; z,C) \prod_{j|p(j)=i}   m_{j\rightarrow i }(x_i)  \hspace{0.3cm},\hspace{0.3cm} x_i\in \mathcal{A} \label{bpmbel}
 \end{equation}

 Since the approximated distribution $\hat{p}(x|y)$ (\ref{appr}) is loop free, the general Belief Propagation theory guarantees that (the normalized) belief vector is
exactly the marginal distribution $\hat{p}(x_i|y)$ of the approximated distribution $\hat{p}(x|y)$ (\ref{appr}).
 To obtain a hard-decision decoding we choose  the symbol whose posterior probability is maximal:
\begin{equation}
\hat{x_i} = \arg \max_a  \mbox{belief}_i(a)  \hspace{1cm},\hspace{1cm} a\in \mathcal{A} \label{bpmhard}
\end{equation}

Above we described the sum-product version of the BP algorithm that computes the marginal probabilities $\hat{p}(x_i|y)$.
A variant of the sum-product algorithm is the max-product algorithm in
which the summation in Eq. (\ref{bpmup})-(\ref{bpmrootdown}) is replaced by a
maximization over all the symbols in $\mathcal{A}$.  The max-product algorithm  finds the most likely pattern of the approximation $\hat{p}(x|y)$.
We did not observe any significant performance difference  using either the sum-product or the max-product variants for decoding MIMO systems presented in the experiment section.
The max-product is more computationally efficient since the BP messages can be entirely computed in the log-domain.

 \comment{We demonstrate empirically in the experiment section that the optimal solution of $\hat{p}(x|y)$ is indeed nearly optimal  for $p(x|y)$.}

\subsection{An MMSE version of a tree approximation}

The MMSE Bayesian approach  (\ref{mmse}) is known to be better than the zero-forcing
 solution (\ref{hzf}). In MMSE we partially incorporate the information that $x\in \mathcal{A}^n$ by using the prior Gaussian distribution $x \sim \mathcal{N} ( 0 , e I)$.
 In a similar way we can  consider a Bayesian version of the proposed Gaussian tree approximation. We can partially incorporate the information that $x\in \mathcal{A}^n$ by using the prior Gaussian distribution $x \sim \mathcal{N} ( 0 , e I)$ such that $e=\frac{1}{|\mathcal{A}|}\sum_{a\in\mathcal{A}}a^2$. This yields the posterior Gaussian distribution:
\begin{eqnarray}
 f_{(x|y)}(x|y)  =
 \frac{1}{\sqrt{ (2 \pi)^n V(x|y)}} \times
 \label{baseap}
\end{eqnarray}
$$ \exp (-\frac{1}{2}(x-E(x|y))^{\T}(\HH^{\T}\! \HH + \frac{\sigma^2}{e}I ) (x-E(x|y))$$
such that $E(x|y) = (\HH^{\T}\! \HH + \frac{\sigma^2}{e}I )^{-1} \HH^{\T}\! y$ and $V(x|y) = (\HH^{\T}\! \HH + \frac{\sigma^2}{e}I )^{-1}$.
The MMSE method is obtained by approximating the unconstrained posterior distribution  $f_{(x|y)}(x|y)$ by a product of marginals. In our approach we use,  instead, the best loop-free Gaussian approximation.
We can apply the Chow-Liu tree approximation on the Gaussian distribution (\ref{baseap}) to obtain a `Bayesian' Gaussian  tree approximation for $p(x|y)$.
This way we partially  use the finite set constraint when we search for the best tree approximation of the true discrete  distribution $p(x|y)$.
This is likely to yield a better approximation of the discrete distribution $p(x|y)$ than the tree distribution which is based on the unconstrained distribution $f(x;z,C)$.

To summarize, our solution to the MIMO decoding  problem is based on applying BP
on a discrete version of the Gaussian tree approximation of the Bayesian version of the continuous least-square solution.
We dub this method   ``The Gaussian-Tree-Approximation (GTA)  Algorithm". The GTA algorithm is   summarized in Fig. \ref{alg_box}.
We next compute the complexity of the GTA algorithm. The complexity of computing the covariance matrix $(\HH^{\T}\HH + \frac{\sigma^2}{e}I)^{-1}$ is $O(n^3)$, the complexity of the Chow-Liu algorithm (based on Prim's algorithm for finding the minimum spanning tree) is $O(n^2)$ and the complexity of the BP algorithm is
$O(|\mathcal{A}|^2 n)$.

\section{Experimental Results}
In this section we provide simulation results for the GTA algorithm over various  MIMO systems.  The channel matrix comprised i.i.d. elements drawn from a zero-mean
complex  normal distribution of unit variance. We  used $500,000$
realizations of  the channel matrix and each matrix was used once for sending a message.  The performance of the proposed algorithm is shown as a
function of the variance of the additive noise $\sigma^2$. The
signal-to-noise ratio (SNR) is defined as $10\log_{10}(E_s/N_0)$
where $E_s/N_0= \frac{n e}{\sigma^2}$ ($n$ is the number of
variables, $\sigma^2$ is the variance of the Gaussian additive noise, and $e$ is the mean symbol energy).
 \begin{figure}[th]
       \centering
         \includegraphics[keepaspectratio=true,clip=true, angle=0,
totalheight=0.25\textheight] {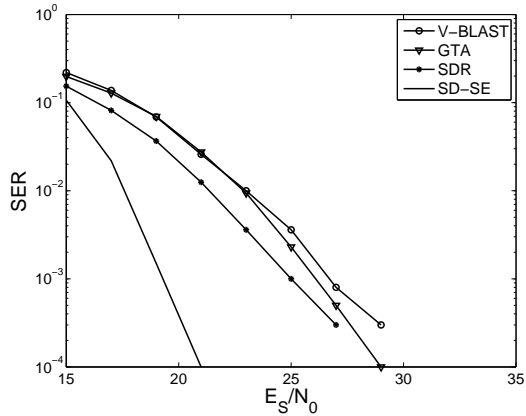} \hspace{1cm}
\caption{Comparison of various detectors in  $12\times 12$ system, 16-QAM symbols. }
\label{fig_12x12cm4}
\end{figure}
We compared the performance of the GTA method to the V-BLAST (MMSE-SIC) algorithm with optimal ordering for the successive interference cancelation \cite{Golden}, and to the Schnorr-Euchner variant of sphere decoding (SE-SD) with infinite radius \cite{agrell2002,Schnorr}.
We used sorting of the channel matrix using the SQRD algorithm \cite{Wubben2001}  and regularization \cite{Wubben2003}, which substantially reduces the computational complexity.

We also implemented the SDR detector suggested by Sidiropoulos and Luo \cite{Sidiropoulos}. Recently it was shown \cite{Ma} that for the cases of 16-QAM and 64-QAM this SDR based method is equivalent to the SDR based detection method suggested by Weisel, Eldar and Shamai \cite{Weisel}.
 The SDRs were solved using the CSDP package \cite{CSDP}. In the SDR Gaussian randomization step, 100 independent randomizations were implemented. All the MIMO detection algorithms were implemented in C for efficiency.

Fig. \ref{fig_12x12cm4} shows MIMO detection performance for a $12 \times 12$ MIMO system using 16-QAM. The methods that are shown are V-BLAST, GTA, SDR and sphere-decoding. In this case the SDR outperforms both the V-BLAST and the GTA methods. However, in this case it is still feasible to compute the optimal maximum-likelihood algorithm using the sphere decoding algorithm.

\begin{figure*}[th]
       \centering
         \includegraphics[keepaspectratio=true,clip=true, angle=0,
totalheight=0.26\textheight] {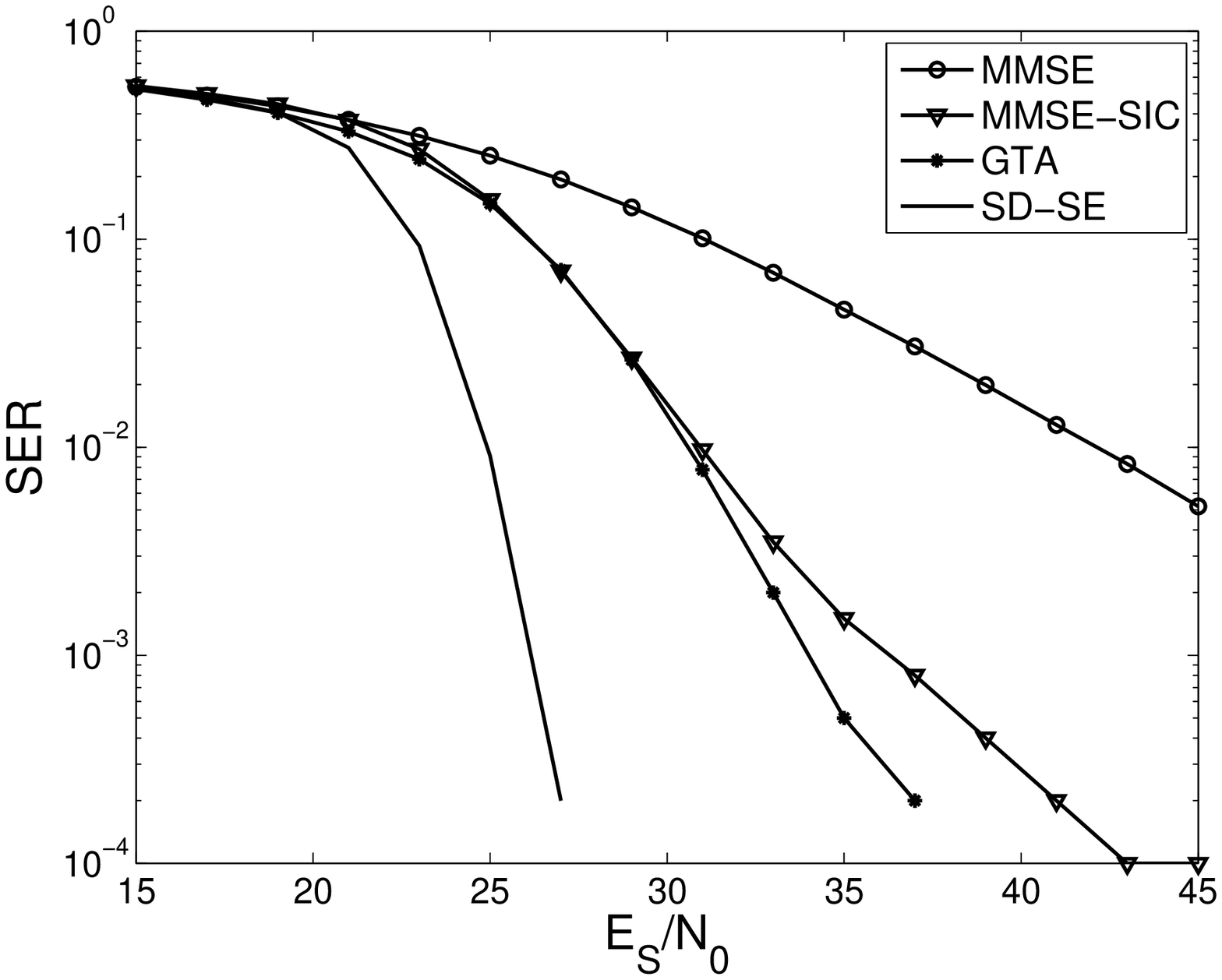} \hspace{1cm}
\includegraphics[keepaspectratio=true,clip=true, angle=0,
totalheight=0.26\textheight] {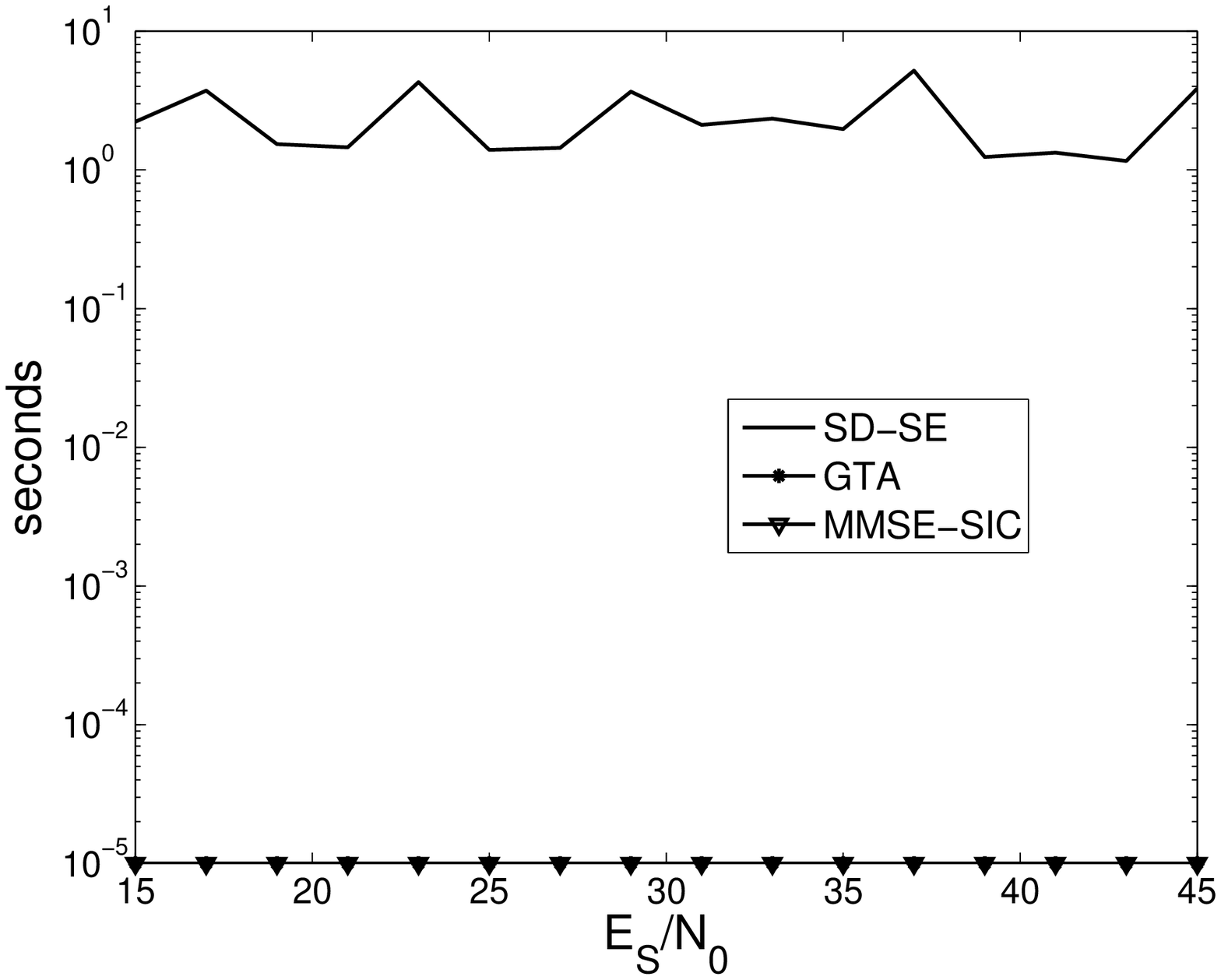}
\centerline{ (a) \hspace{7cm} (b)}
\caption{ $12\times 12$ system, 64-QAM symbols. (a) SER versus SNR, (b) max seconds for decoded symbol vector versus SNR. }
\label{fig_8x8}
\end{figure*}

\begin{figure*}[th]
       \centering
         \includegraphics[keepaspectratio=true,clip=true, angle=0,
totalheight=0.26\textheight] {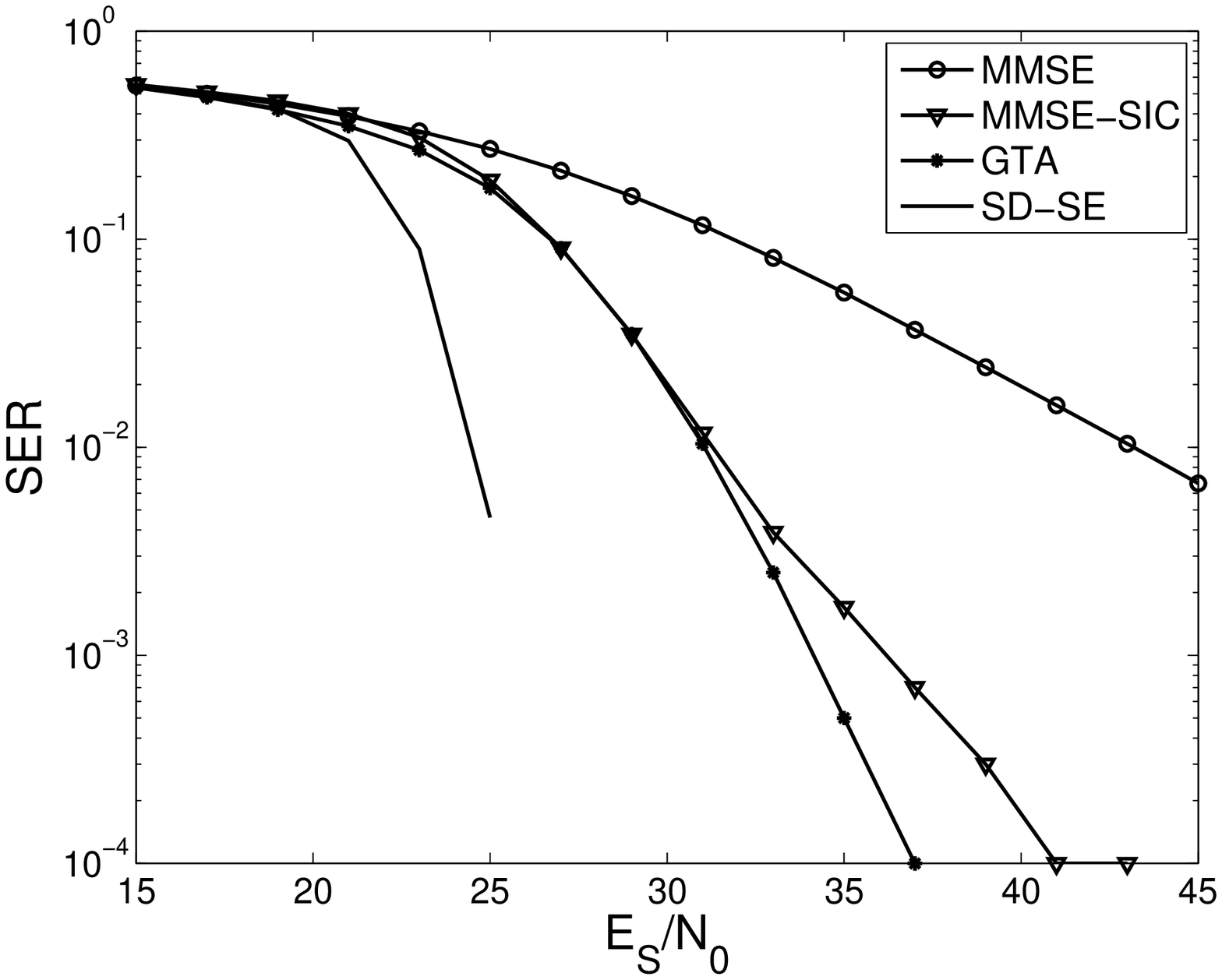}\hspace{1cm}
\includegraphics[keepaspectratio=true,clip=true, angle=0,
totalheight=0.26\textheight] {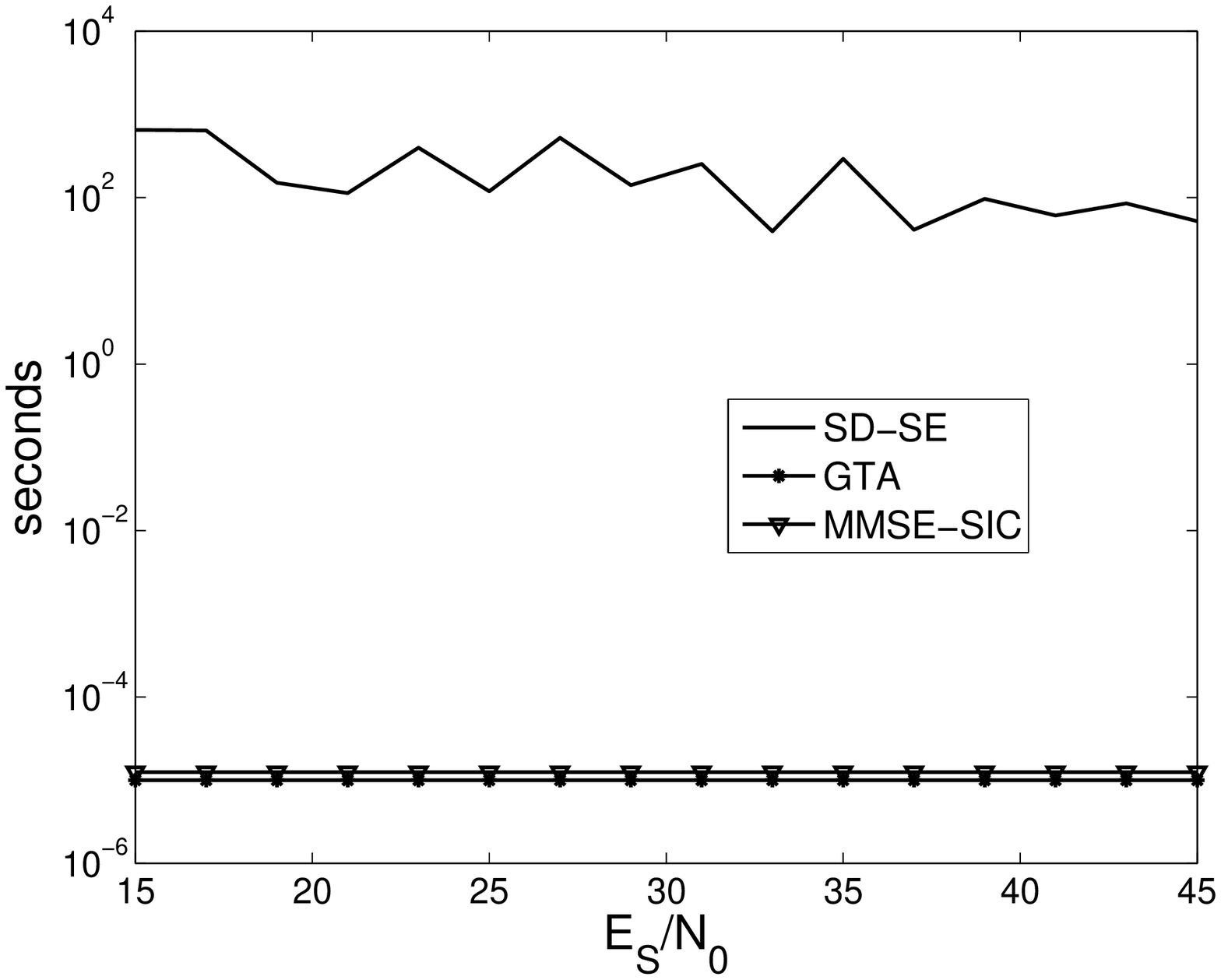}
\centerline{ (a) \hspace{7cm} (b)}
\caption{ $16\times 16$ system, 64-QAM symbols. (a) SER versus SNR, (b) max seconds for decoded symbol vector versus SNR. }
\label{fig_12x12}
\end{figure*}

The SD-SE is the favorite detection method when the problem size is small or moderate. In this case the SD-SE can always yield the exact ML solution at acceptable computational cost.
In large size problems, however, there is still a need for good approximation methods.
Fig. \ref{fig_8x8} shows the SER versus SNR and worst case execution time versus SNR for the $12\times 12$ system using 64-QAM. Fig. \ref{fig_12x12} shows the same experimental results for the
$16\times 16$ system using 64-QAM.
To assess the computational complexity we used a measure of the worst case rather than the average case since in online applications we have to decode within a specified time. The choice between execution time or number of floating point operations is debatable. The differences in running time between the methods we implemented was in orders of magnitude and running time is easier to appreciate.

 As can be seen from Figs. \ref{fig_8x8} and \ref{fig_12x12}, the performance of the GTA algorithm in high SNR is significantly better than the V-BLAST. The computational complexity of GTA is comparable to V-BLAST and it is two orders of magnitude better than SDR.
 We note in passing that the performance of the SDR method \cite{Sidiropoulos} in these 64-QAM cases is worse than that  of MMSE-SIC and the computational complexity is much higher.
 From the derivation of the SDR method it can be seen that the relaxation becomes more crude as at higher constellations.

\begin{figure}[th]
\center
          \includegraphics[keepaspectratio=true,clip=true, angle=0,
totalheight=0.28\textheight] {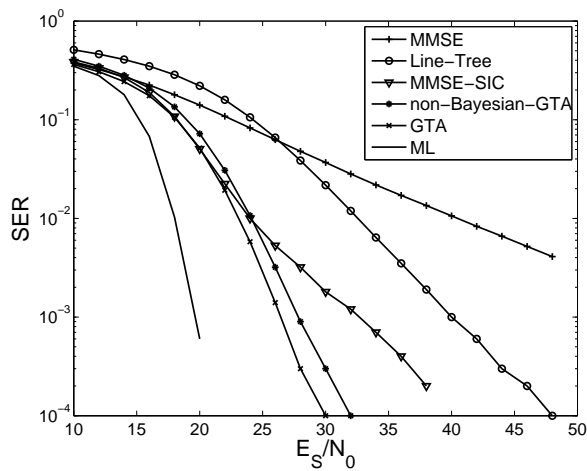}
 \caption{Comparative results of MMSE, MMSE-SIC and variants of the GTA for $20 \times 20$ real system ,$\mathcal{A}=\{\pm1,\pm3\}$.
}
\label{fig_res}
\end{figure}

We next show the performance of several variants of the GTA algorithm.
 The GTA algorithm differs from the ZF, MMSE and MMSE-SIC algorithms in several ways. The first is a Markovian approximation of $f(x;z,C)$ instead of an approximation based on a product of independent densities. The second aspect is  the use of an optimal tree. To clarify the contribution of each component we modified the GTA algorithm by replacing the Chow-Liu optimal tree by the tree $1 \rightarrow 2  \rightarrow 3,...,\rightarrow n$. We call this method the `Line-Tree'. As can be seen from Fig. \ref{fig_res}, using the optimal tree is crucial to obtain improved results. Fig. \ref{fig_res} also shows  the results of the non-Bayesian variant of the GTA algorithm. As can be seen, the Bayesian version yields  better results. Fig. \ref{fig_res} shows the symbol error rate (SER) versus SNR for a
 $20\!\times\! 20$, $|\mathcal{A}|=4$, real MIMO system.
The performance of the  GTA method and its variants was compared to
the MMSE and the MMSE-SIC algorithms.

\section{Conclusion}

  We proposed a novel
 MIMO detection technique based on the principle of a tree approximation of the Gaussian distribution that corresponds to the continuous linear problem.
 The proposed method outperforms previously suggested MIMO decoding  algorithms in high order QAM constellation, as demonstrated in simulations.
 Although the proposed method yields improved results, the tree approximation we applied may not be the best one (finding the best tree for the integer constrained linear problem is NP hard). It is left for future research to search for a better discrete tree approximation for the constrained linear least squares problem.  This paper dealt with a tree approximation approach, more complicated approximations such as multi-parent trees could improve performance and can potentially provide a smooth performance-complexity trade-off.

While the method provides excellent performance, it is worthwhile
mentioning  that the method provides a-posteriori probabilities for
each variable that can be used to improve performance. This is done
by applying a Schnorr-Euchner sphere decoding algorithm
\cite{Schnorr}, where we order the symbols according to their
a-posteriori probabilities. This can lead to very
close  to optimal performance in a significantly reduced complexity.
It is because that by computing the a-posteriori probabilities, we
have a much higher probability of finding the true solution during
the first search, therefore significantly reducing the search
radius.

There are several important applications of the proposed technique.
We comment here on combining it into communication systems with
coding and interleaving. It is useful both for single carrier and
OFDM systems.  It can serve as a MIMO decoder
for wireless communication systems. Using the a-posteriori
probability distribution of the symbols we can easily compute the
a-posteriori probability and the likelihood ratio for the bits.
  The technique can be combined with MIMO-OFDM system with
bit interleaved coded modulation or with trellis coded modulation by
joint coding of over all frequency tones of the OFDM system, and
running the decoder for each tone independently.

In this paper we focused on the MIMO detection problem. The proposed method, however, can be applied to solve constrained linear  least squares problems
which is an important issue in many fields.
A main concept in the GTA model is the interplay between discrete and Gaussian models.
Such hybrid ideas can be considered also for discrete inference problems other than least-squares.

\bibliographystyle{plain}

\begin{thebibliography}{10}

\bibitem{agrell2002}
E.~Agrell, T.~Eriksson, A.~Vardy, and K.~Zeger.
\newblock Closest point search in lattices.
\newblock {\em IEEE Transactions on Information Theory}, 48(8):2201--2214,
  2002.

\bibitem{Bayati}
M.~Bayati, D.~Shah, and M.~Sharma.
\newblock Maximum weight matching via max-product belief propagation.
\newblock {\em ISIT}, 2005.

\bibitem{Bickson}
D.~Bickson, O.~Shental, P.~H. Siegel, J.~K. Wolf, and D.~Dolev.
\newblock Gaussian belief propagation based multiuser detection.
\newblock {\em ISIT}, 2008.

\bibitem{CSDP}
B.~Borchers.
\newblock A {C} library for semidefinite programming.
\newblock {\em Optimization Methods and Software}, 11:613--623, 1999.

\bibitem{Boutros}
J.~Boutros, N.~Gresset, L.~Brunel, and M.~Fossorier.
\newblock Soft-input soft-output lattice sphere decoder for linear channels.
\newblock {\em GLOBECOM}, 2003.

\bibitem{Chow}
C.~K. Chow and C.~N. Liu.
\newblock Approximating discrete probability distributions with dependence
  trees.
\newblock {\em IEEE Trans. on Info. Theory}, pages 462--467, 1968.

\bibitem{Felzenszwalb}
P.~F. Felzenszwalb and D.~P. Huttenlocher.
\newblock Efficient belief propagation for early vision.
\newblock {\em International Journal of Computer Vision}, pages 41--54, 2006.

\bibitem{Fincke}
U.~Fincke and M.~Pohst.
\newblock Improved methods for calculating vectors of short length in a
  lattice, including a complexity analysis.
\newblock {\em Math. Computat.}, pages 463--471, 1985.

\bibitem{Gallager}
R.~G. Gallager.
\newblock Low density parity check codes.
\newblock {\em IRE Trans. Inform.Theory}, pages 21--28, 1962.

\bibitem{Goldberger_NIPS}
J.~Goldberger and A.~Leshem.
\newblock A {G}aussian tree approximation for integer least-squares.
\newblock {\em Neural Information Processing Systems 23 (NIPS)}, 2009.

\bibitem{Golden}
G.~D. Golden, G.~J. Foschini, , R.~A. Valenzuela, and P.~W. Wolniansky.
\newblock Detection algorithm and initial laboratory results using {V-BLAST}
  space-time communication architecture.
\newblock {\em Electron. Letters}, 35:14--16, 1999.

\bibitem{Hu}
J.~Hu and T.~M. Duman.
\newblock Graph-based detector for {BLAST} architecture.
\newblock {\em IEEE Communications Society ICC}, 2007.

\bibitem{Jalden}
J.~Jalden and B.~Ottersten.
\newblock An exponential lower bound on the expected complexity of sphere
  decoding.
\newblock {\em IEEE Intl. Conf. Acoustic, Speech, Signal Processing}, 2004.

\bibitem{xJalden2005}
J.~Jalden and B.~Ottersten.
\newblock On the complexity of sphere decoding in digital communications.
\newblock {\em IEEE Trans. Signal Processing}, pages 1474--1484, 2005.

\bibitem{Kabashima}
Y.~Kabashima.
\newblock A {CDMA} multiuser detection algorithm on the basis of belief
  propagation.
\newblock {\em J. Phys. A}, pages 11111--11121, 2003.

\bibitem{Kaynak}
M.~Kaynak, T.~Duman, and E.~Kurtas.
\newblock Belief propagation over {SISO/MIMO} frequency selective fading
  channels.
\newblock {\em IEEE Transactions on Wireless Communications}, pages 2001--5,
  2007.

\bibitem{Kschischang}
F.~Kschischang, B.~J. Frey, and H.~Loeliger.
\newblock Factor graphs and the sum-product algorithm.
\newblock {\em IEEE Transactions on Information Theory}, 47:498--519, 2001.

\bibitem{Ma}
W.K. Ma, C.C. Su, J.~Jalden, T.H. Chang, and C.Y. Chi.
\newblock The equivalence of semidefinite relaxation {MIMO} detectors for
  higher-order {QAM}.
\newblock {\em IEEE Journal of Selected Topics of Signal Processing}, in press.

\bibitem{Minka}
T.~Minka and Y.~Qi.
\newblock Tree-structured approximations by expectation propagation.
\newblock {\em Advances in Neural Information Processing Systems}, 2004.

\bibitem{Montanari}
A.~Montanari, B.~Prabhakar, and D.~Tse.
\newblock Belief propagation based multi-user detection.
\newblock {\em Allerton Conference on Communication, Control and Computing},
  2005.

\bibitem{Murphy}
K.~Murphy, Y.~Weiss, and M.~Jordan.
\newblock Loopy belief propagation for approximate inference: An empirical
  study.
\newblock {\em Uncertainty in Artificial Intelligence (UAI)}, pages 467--475,
  1999.

\bibitem{Neirotti}
J.P. Neirotti and D.~Saad.
\newblock Improved message passing for inference in densely connected systems.
\newblock {\em Europhys. Lett.}, pages 866--872, 2005.

\bibitem{Pearl}
J.~Pearl.
\newblock Probabilistic reasoning in intelligent systems.
\newblock {\em San Mateo CA: Morgan Kaufman}, 1988.

\bibitem{Prim}
R.~C. Prim.
\newblock Shortest connection networks and some generalizations.
\newblock {\em J. Bell System Tech.}, pages 1389--1401, 1957.

\bibitem{Schnorr}
C.P. Schnorr and M.~Euchner.
\newblock Lattice basis reduction: {I}mproved practical algorithms and solving
  subset sum problems.
\newblock {\em Journal of Mathematical Programming}, 66(1-3):181--199, 1994.

\bibitem{Shental2008}
O.~Shental, N.~Shental, S.~Shamai (Shitz), I.~Kanter, A.J. Weiss, and Y.~Weiss.
\newblock Discrete-input two-dimensional {G}aussian channels with memory:
  Estimation and information rates via graphical models and statistical
  mechanics.
\newblock {\em Information Theory, IEEE Transactions on}, pages 1500--1513,
  2008.

\bibitem{Sidiropoulos}
N.D. Sidiropoulos and Z.Q. Luo.
\newblock A semidefinite relaxation approach to {MIMO} detection for high-order
  {QAM} constellations.
\newblock {\em IEEE Signal Processing Letters}, 13(9):525--528, 2006.

\bibitem{Studer}
C.~Studer, A.~Burg, and H.~Bolcskei.
\newblock Soft-output sphere decoding: Algorithms and {VLSI} implementation.
\newblock {\em IEEE Journal on Selected Areas in Communications}, pages
  290--300, 2008.

\bibitem{Sudderth}
E.~Sudderth and W.~Freeman.
\newblock Signal and image processing with belief propagation.
\newblock {\em IEEE Signal Processing Magazine, DSP Applications Column}, pages
  114--121, 2008.

\bibitem{Weiss2001}
Y.~Weiss and W.T. Freeman.
\newblock Correctness of belief propagation in {Gaussian} graphical models of
  arbitrary topology.
\newblock {\em Neural Computation}, pages 2173--2200, 2001.

\bibitem{Weisel}
A.~Wiesel, Y.~C. Eldar, and S.~Shamai.
\newblock Semidefinite relaxation for detection of 16-{QAM} signaling in {MIMO}
  channels.
\newblock {\em IEEE Signal Processing Letters}, 2005.

\bibitem{Wubben2003}
D.~W\"{u}bben, R.~B\"{o}hnke, V.~K\"{u}hn, and K.~Kammeyer.
\newblock {MMSE} extension of {V-BLAST} based on sorted {QR} decomposition.
\newblock {\em IEEE Proc. Vehicular Technology Conference}, 1:508--512, 2003.

\bibitem{Wubben2001}
D.~W\"{u}bben, R.~B\"{o}hnke, J.~Rinas, V.~K\"{u}hn, and K.~Kammeyer.
\newblock Efficient algorithm for decoding layered space-time codes.
\newblock {\em IEE Electronics Letters}, 37:1348--1350, 2001.

\bibitem{Yedidia}
J.~S. Yedidia, W.~T. Freeman, and Y.~Weiss.
\newblock Understanding belief propagation and its generalizations.
\newblock {\em IJCAI}, 2001.

\end{thebibliography}

\end{document}